# Using Dynamic Allocation of Write Voltage to Extend Flash Memory Lifetime

Haobo Wang, *Student Member, IEEE*, Nathan Wong, *Student Member, IEEE*, Tsung-Yi Chen, *Member, IEEE*, and Richard D. Wesel, *Senior Member, IEEE*

*Abstract*—The read channel of a Flash memory cell degrades after repetitive program and erase (P/E) operations. This degradation is often modeled as a function of the number of P/E cycles. In contrast, this paper models the degradation as a function of the cumulative effect of the charge written and erased from the cell. Based on this modeling approach, this paper dynamically allocates voltage using lower-voltage write thresholds at the beginning of the device lifetime and increasing the thresholds as needed to maintain the mutual information of the read channel in the face of degradation. The paper introduces the technique in an idealized setting and then removes ideal assumptions about channel knowledge and available voltage resolution to conclude with a practical scheme with performance close to that of the idealized setting.

## I. INTRODUCTION

FLASH memory has been widely employed in both consumer electronic devices and industrial electronic systems because of its ability to support high-throughput and low-latency memory access. However, a fundamental issue with Flash technology is that its read channel experiences significant degradation over time which eventually produces unacceptable reliability. As modern Flash solutions provide more storage capacity in smaller form factors, the resulting increase of physical cell density and signal constellation density amplifies the degradation problem.

The degradation can be addressed in different layers in the Flash system. At the device layer three-dimensional cell structures improve durability [1]–[3]. At the system level channel codes such as BCH codes [4]–[6] and more recently LDPC codes [7]–[10] add redundancy to protect the stored information. In [11], the authors write to cells with a lower voltage but for a longer time to cause less damage at the expense of increased write time. The resulting scheme provides lifetime extension while still guaranteeing a desired write throughput.

The degradation over time is often modeled as a function of the number of program and erase (P/E) cycles, so a direct

This material is based upon work supported partially by Western Digital Inc. Micron Technology Inc. through the UCLA Center on the Development of Emerging Storage Systems, and the National Science Foundation under Grant Number 1162501. Any opinions findings, and conclusions or recommendations expressed in this material are those of the author(s) and do not necessarily reflect the views of the National Science Foundation.
H. Wang, N. Wong and R. D. Wesel are with the Department of Electrical Engineering, University of California, Los Angeles, CA 90095 USA (e-mail: whb12@ucla.edu; nsc.wong@ucla.edu; wesel@ee.ucla.edu).
T.-Y. Chen is with SpiderCloud Wireless Inc., San Jose, CA 95134 USA (e-mail: steven.tychen@gmail.com).

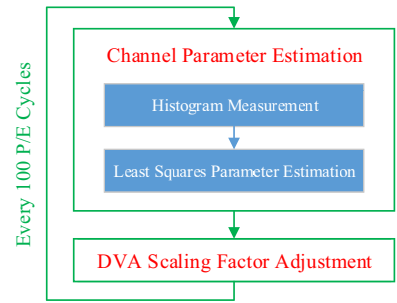

Fig. 1. Estimation-based DVA framework.

solution is reducing the number of P/E cycles needed. Write-once memory (WOM) codes [12]–[15] provide one approach to reduce the number of P/E cycles required to store information by permitting multiple writes before an erase cycle is needed. Another way to reduce the number of P/E cycles is rank modulation [16]–[18], which stores information in the cell using the relative value (or ordering) of cell charge levels rather than the absolute value. Thus a block of cells can be rewritten without erasing by adding charge to properly re-order the cells.

Several papers have explored dynamically adjusting to the degrading read channel [19]–[21]. In [19], read thresholds are progressively adjusted to minimize hard decoding BER or provide better log-likelihood for soft decoding based on previous reads. In [20], [21], dynamic threshold assignment (DTA) adjusts the read thresholds to match the shifting and widening threshold voltage distributions of the read channel, significantly improving bit error rate (BER) performance.

In contrast to the P/E-cycle-based degradation model, this paper models the degradation as a function of the cumulative effect of the charge written and erased from the cell, which we call the accumulated voltage $V_{acc}$. The accumulated voltage model reveals the opportunity to improve lifetime by minimizing the $V_{acc}$ required to store a given amount of information. In particular, this paper explores dynamically adjusting the target threshold voltage levels by using lower target threshold voltage levels at the beginning of the device lifetime. This approach is called dynamic voltage allocation (DVA). As the read channel becomes more degraded, the threshold voltages are gradually increased to what would be the nominal values in a standard device not employing DVA. As shown in our precursor conference papers [22], [23], DVA maintains the required mutual information while significantly extending lifetime.

Fig. 1 illustrates the basic structure of the DVA approach. Periodically (every 100 P/E cycles in the figure) multiple reads using different read thresholds produce a histogram of the threshold voltages of the cells on the page or pages considered. A parameter-based least-squares channel estimation determines the quality of the read channel from this histogram. Based on the channel estimation, write threshold voltages are set to be as low as possible while still ensuring that the read channel has sufficient mutual information to be successfully decoded.

This paper systematically explains and demonstrates the concept of dynamic voltage allocation, and addresses several issues to support a practical implementation. This paper extends our previous work [22], [23] in the following ways:

1) Adding programming error and cell-to-cell interference to the ground truth channel model;
2) Analyzing DVA's performance when both estimation and scale-factor adjustment algorithms are simplified by using a simple Gaussian model to approximate the underlying (and more complex) ground truth channel;
3) Designing and analyzing DVA specialized to match an even-odd structure for writing to cells;
4) Exploring the improvement obtained by DVA over DTA;
5) Analyzing DVA's performance when both write and read voltage values are restricted to a finite set of available voltages;
6) Analyzing DVA's complexity.

This paper begins by introducing DVA in an idealized setting and then removes ideal assumptions about channel knowledge and threshold resolution to conclude with a practical scheme that has performance similar to that of the idealized setting. DVA is first explored under the assumption of perfect knowledge of the channel state as in our precursor conference paper [22], except with a more complex read channel that includes cell-to-cell interference and programming errors.

Then the performance of DVA is explored when it must gain its channel information through estimation. All dynamic schemes such as DTA and DVA require some form of information about the read channel, and different methods can be employed to acquire the information. For DTA, knowledge of the voltage distribution is needed. In [20], repetitive read operations are needed at relatively precise voltages to enable a bisection algorithm to place the read thresholds. In [21], threshold measurements of a certain number of cells are required.

As shown in Fig. 1, our approach to acquire read channel information is to measure a limited-resolution histogram and interpret this histogram using certain assumptions about the channel model. In [24], the authors model the Flash read channel as a multi-modal Gaussian distribution with means and variances as parameters, and demonstrate that least squares algorithms estimate the means and variances well with histograms having as few as twelve bins. These estimated parameters are used to set read thresholds according to [25].

This channel estimation is first performed with a perfect channel model as explored in our precursor conference paper [23]. Next, this paper explores the practical scenario in which the channel model does not perfectly represent the true channel. This mismatch is both a reflection of the imperfect characterization information available about a specific Flash device and the fact that, even with perfect knowledge of the channel model, simpler models might be preferable because they reduce the complexity of estimation.

Next, we explore how constraining the resolution of write and read thresholds affects performance. We also compare the performance of DVA to DTA and analyze complexity of the DVA framework. Possible solutions to reduce the cost of implementing DVA are proposed. Throughout the paper, Multi-level Cell (MLC) Flash (with four levels) is assumed for all the models and simulations.

The remainder of this paper is organized as follows: Sec. II presents the complete channel model. Simplified models are used in some sections. Sec. III introduces DVA using the complete channel model but in the idealized setting of perfect channel state information. Sec. IV formulates the channel parameter estimation problem and presents the least squares algorithm and binning strategy used in this paper. Sec. V examines the practical scenario in which channel estimation and DVA are based on a model that is not perfect because it is simpler than the actual channel. Sec. VI compares the performance of DTA and DVA. Sec. VII adds practical constraints on the resolution of the read and write threshold voltages. Sec. VIII analyzes the complexity cost of DVA framework. Sec. IX concludes the paper.

## II. MODELING CHANNEL PARAMETERS & DEGRADATION

Because the interfaces are proprietary, we are not able to measure data from actual flash devices. Instead, we use the models introduced in this section to generate the noise that reflects the behavior of the Flash memory read channel. These models are called *ground truth models*, which means that they are used for all simulations. The term "ground truth model" distinguishes these models from less precise simple Gaussian model (which is called the *channel model assumption*) that are used by the channel estimation and DVA algorithms. Note that all algorithms, regardless of the models they incorporate in their calculations, are simulated on the ground truth models.

The ground truth models are not matched to a particular Flash device, but based on the academic publications we cite as we present the models. We believe these models reflect the major channel degradations and provide a reasonable degradation trajectory over the lifetime of Flash memory. We use these qualitatively correct channel models to show that DVA can counter the major types of degradation in Flash memory channels. The models in this section can be replaced with device-specific models to apply DVA to a particular Flash memory system. In fact, we will show in Sec. V that DVA does not need a precise channel model to provide a significant improvement in lifetime.

Based on [26]–[37], our precursor conference papers [22] and [23] propose a parameterized channel model characterizing the noise as having three additive components: programming noise, wear-out noise, and retention noise. This paper improves the model of [22] and [23] by adding the effects of cell-to-cell interference and programming error.

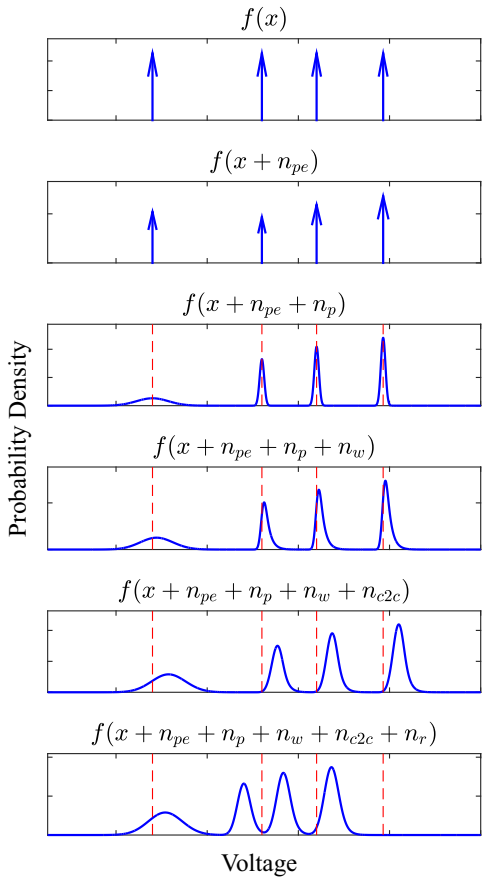

Fig. 2. Flash read channel PDFs illustrating how the probability density of voltage thresholds is affected by various noise components.

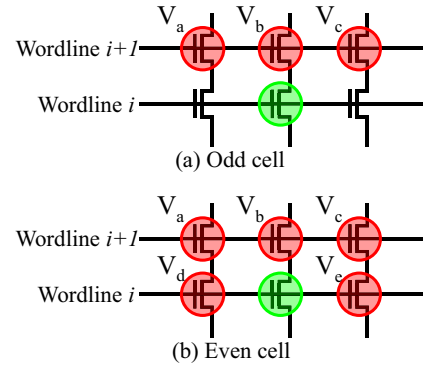

Fig. 3. Spacial relationship between the interfering cells (red) and interfered cells (green). Flash memory is written wordline by wordline, i.e., wordline $i+1$ is written after wordline $i$. For the interference shown above, within a wordline even cells are written before odd cells.

### A. Channel Model with Additive Components

We formulate a Flash memory read channel model with five additive noise components as follows:

$$y = x + n_{pe} + n_p + n_w + n_{c2c} + n_r, \quad (1)$$

where voltage $x$ is the intended threshold voltage written to a cell, and $y$ is the measured threshold voltage. Noise $n_p$ denotes the programming noise, $n_w$ denotes the wear-out noise, $n_r$ denotes the retention noise, $n_{c2c}$ denotes the cell-to-cell interference, and $n_{pe}$ denotes the programming error.

Fig. 2 shows an example of voltage distribution probability distribution functions (PDFs) which demonstrates the additive effect of each noise component. The arrows in the figure represent delta functions.

*1) Programming Error $n_{pe}$ [38]:* Programming errors occur when a bit of the lower page is mis-read in preparation for writing a bit of the upper page to an MLC cell. Essentially, as the bit of the upper page is written, the error of the first bit is amplified. The programming error is modeled with a probability mass function (PMF). For an intended level $x$, channel parameter $P_{x,y} = P(Y = y|X = x)$ is the conditional probability of actually writing $y$. For MLC Flash, we have

$$P_{x,0} + P_{x,1} + P_{x,2} + P_{x,3} = 1, \quad (2)$$

for each $x \in \{0, 1, 2, 3\}$. These channel parameters are strongly related to the number of P/E cycles.

This distortion changes the distribution of stored data. For example, if there is no other noise component and the original data is uniformly distributed (i.e. each level is equally likely), programming errors move some write thresholds to higher levels so that the levels are no longer equally likely. Results in [38] indicate that the impact of programming errors becomes significant only after a large number of P/E cycles.

*2) Programming Noise $n_p$ [22], [23], [26], [27]:* The programming noise is modeled as a Gaussian noise for each level, and the noise variance of programmed states is smaller than that of the erased state because the feedback control loop associated with programming reduces threshold variation. The PDF of the programming noise is represented as

$$f_{n_p}(n_p|x = l) = \begin{cases} \mathcal{N}(0, \sigma_e^2) & \text{if } l = 0 \\ \mathcal{N}(0, \sigma_p^2) & \text{if } l > 0 \end{cases}, \quad (3)$$

where $\sigma_e > \sigma_p$. Index $l$ represents the level of the intended threshold voltage level. For MLC Flash, $l \in \{0, 1, 2, 3\}$ where $l = 0$ indicates the erased state. Standard deviations $\sigma_e$ and $\sigma_p$ are the channel parameters for this component, and remain constant throughout the lifetime of the device.

*3) Wear-out Noise $n_w$ [22], [23], [28]–[33]:* The wear-out noise is modeled as a positive-side exponential[1] noise for each level, and the slope of the distribution is characterized by the channel parameter $\lambda$. The PDF is

$$f_{n_w}(n_w) = \begin{cases} \frac{1}{\lambda} e^{-\frac{n_w}{\lambda}} & \text{if } n_w \geq 0, \\ 0 & \text{if } n_w < 0. \end{cases} \quad (4)$$

Parameter $\lambda$ increases as the device experiences more P/E cycles (or larger $V_{acc}$).

*4) Cell-to-cell Interference $n_{c2c}$ [37], [39]:* The cell-to-cell interference experienced by a cell is a weighted sum of the voltage increases in neighboring cells that occur *after* the cell of interest has been written [40]. Fig. 3 shows the spacial relationships between the interfering cells and the cell

---

[1]For various device implementations wear-out noise can also be a negative-side exponential or a double-sided exponential (Laplace) distribution. The DVA and channel estimation techniques we present can be applied in all these cases.

$$\mu_r = -(V - V_0) \ln\left(1 + \frac{t}{t_0}\right) \left[A_r \left(\frac{V_{acc}}{V_{max}}\right)^{k_i} + B_r \left(\frac{V_{acc}}{V_{max}}\right)^{k_o}\right] \quad (5)$$

$$\sigma_r^2 = 0.1(V - V_0) \ln\left(1 + \frac{t}{t_0}\right) \left[A_r \left(\frac{V_{acc}}{V_{max}}\right)^{k_i} + B_r \left(\frac{V_{acc}}{V_{max}}\right)^{k_o}\right]^2 \quad (6)$$

of interest for Flash memories employing the common even-odd structure for writing and for reading. The green circle indicates the cell of interest, and the red circles indicate the interfering cells. Fig. 3(b) shows that even cells, which are written first, suffer interference from the odd cells in that wordline, which are written after the even cells. As is shown in Fig. 3(a), the odd cells do not suffer any interference from the even cells in the same wordline because the odd cells are written after the even cells. All the cells in wordline $i$ suffer interference from the neighboring cells in wordline $i+1$, which is written after wordline $i$. The threshold voltage disturbance (increase) of odd cells $V_{c2c,odd}$ and even cells $V_{c2c,even}$ caused by this interference can be modeled as

$$V_{n_{c2c,odd}} = \gamma_a V_a + \gamma_b V_b + \gamma_c V_c \quad (7a)$$
$$V_{n_{c2c,even}} = V_{n_{c2c,odd}} + \gamma_d V_d + \gamma_e V_e. \quad (7b)$$

As shown in Fig. 3, $V_a$, $V_b$, and $V_c$ are voltage increases from the cells in the next wordline. $V_b$ is from the cell directly above the cell of interest. $V_a$ and $V_c$ are voltage increases in cells diagonally adjacent on the next wordline that are located either upper left ($V_a$) or upper right ($V_c$). The voltage increases $V_d$ and $V_e$ are from adjacent cells on the same wordline that are to the left ($V_d$) or right ($V_e$). When these cells are programmed (subsequent to the programming of the cell of interest), they interfere with the cell of interest according to the coupling factors ($\gamma$'s) between the interfering cells and the cell of interest. The magnitude of this noise component is thus related to two factors: the values of the voltage increases as the adjacent cells are written to their intended thresholds and the coupling factors $\gamma$.

*5) Retention Noise $n_r$ [22], [23], [29], [34]–[37]:* The retention noise is modeled as a Gaussian random variable with PDF

$$f_{n_r}(n_r) = \frac{1}{\sigma_r \sqrt{2\pi}} e^{-\frac{(n_r - \mu_r)^2}{2\sigma_r^2}}. \quad (8)$$

Both $\mu_r$ and $\sigma_r^2$ (characterized in detail below in Sec. II-B) are dynamic channel parameters determined by the number of P/E cycles (or $V_{acc}$), retention time and intended threshold voltage.

### B. Channel Parameter Degradation Model

This subsection presents a model describing how the parameters describing the additive noise terms degrade as a function of P/E operations and retention time. From the discussion in Sec. II-A, this degradation occurs for all the channel parameters except those describing the programming noise $n_p$.

*1) Degradation model due to P/E operations [22], [23]:* The degree of damage to a Flash cell's oxide layer caused by P/E operations is directly related to the volume of charge traveling through the oxide layer [41]. While the channel degradation caused by P/E operations is often modeled as a function of the number of P/E cycles, this model essentially assumes that approximately the same volume of charge travels through the oxide layer during each P/E cycle. The novel approach of DVA is to realize that the intended threshold voltages themselves can be varied over time so that less charge travels through the oxide layer during early P/E cycles when the channel is favorable. Because it is the charge stored in the floating gate that changes the threshold voltage, the difference between the intended threshold voltage and the erased state voltage is a good indicator of the amount of charged transferred. Thus, we define the voltage-based metric $V_{acc}/V_{max}$ to characterize P/E cycling where $V_{acc}$ denotes the accumulated voltage over P/E cycles

$$V_{acc} = \sum_{j=1}^{N}(V_p^{(j)} - V_e), \quad (9)$$

and $V_{max}$ is the maximum voltage difference between programmed and erased states of a Flash cell. In (9), $N$ is the number of P/E cycles, $V_p^{(j)}$ is the intended threshold in $j$th P/E operation, and $V_e$ is the intended threshold voltage of the erased state. Only the programming process is considered in this model, because program and erase are symmetric operations from the perspective of the amount of charge passing through the floating gate.

*2) Degradation model for the wear-out noise parameter [30], [33]:* P/E cycling operations cause the formation of oxide traps and interface traps in the cells [28], [29], [31], [32]. In [35] a power law describes how interface trap density depends on P/E cycle count as follows: $A_w \cdot$ (P/E cycle count)$^{k_i}$, where $k_i$ and $A_w$ are constant degradation parameters determined by the underlying physical properties of the Flash device. Replacing the P/E cycle count with $V_{acc}/V_{max}$ yields our expression for interface trap density: $A_w \cdot ((V_{acc}/V_{max})^{k_i}$. Based on this expression and [30], [33], the wear-out channel parameter $\lambda$ in (4) has a degradation model that can be formulated as

$$\lambda = C_w + A_w \cdot \left(\frac{V_{acc}}{V_{max}}\right)^{k_i}, \quad (10)$$

where $C_w$ is another constant degradation parameters in addition to $k_i$ and $A_w$.

*3) Degradation model for the retention noise parameter [29], [35], [37]:* Retention noise models channel degradation in the form a gradual decrease of threshold voltage.

Both trap recovery and electron detrapping contributes to this degradation [34]–[36]. In addition to the interface trap density model, [35] suggests that oxide trap density as well can be modeled by a power of the P/E cycle count. With the $V_{acc}$-based characterization of P/E cycling, total trap density can be represented as $A_r \cdot (V_{acc}/V_{max})^{k_i} + B_r \cdot (V_{acc}/V_{max})^{k_o}$. From [29], [35], [37], we define the degradation model for the retention noise channel parameters $\mu_r$ and $\sigma_r^2$ of (4) as (5) and (6) [22], where voltage $V$ is the intended threshold voltage, and $V_0$ is the erased state threshold voltage. $k_i, k_o, A_r$ and $B_r$ are constant degradation parameters determined by the physical properties of individual Flash system. Parameter $t$ is the retention time and $t_0$ is its normalization factor. In this paper, we study Flash read channel characteristics at a fixed retention target of one year ($t = 8760$ hours). The techniques presented in this paper can, of course, be applied for any target retention time.

*4) Cell-to-cell Interference Parameter Model:* In [39], the $\gamma$ parameters in (8) are modeled as random variables with truncated Gaussian distributions. The authors of [39] mention that the means of interfering cells' $\gamma$'s in the next wordline to be programmed are also random. We simplify the model by assuming the means of all the $\gamma$'s are constant. In our research, the parameter model provided in [39] is utilized, and the cell-to-cell interference strength factor is set to $s = 0.2$ which yields the full set of cell-to-cell interference parameters given in the appendix. This setting generates a pair of even-cell and odd-cell read channels with sufficient difference and reasonable typicality.

*5) Programming Error Parameter Degradation Model [38]:* We employ the programming error parameter degradation model proposed in [38]. $P_{X,Y}$ has an exponential relationship with P/E cycle counts.

$$P_{X,Y} = \exp(c_1 x + c_0) \quad (11)$$

The parameters $c_1$ and $c_0$ are different for each $P_{X,Y}$, and variable $x$ indicates the *normalized* P/E cycle count where the manufacture specified lifetime is used as the normalization factor. In this paper, the normalization factor is set to 3000 P/E cycles. The resulting $P_{X,Y}$ expressions for our model are given in the appendix.

### C. Models Used in This Paper

The results presented in this paper are software simulation results based on two channel models that we have constructed based on the literature. Model 1 is the channel model used in [23], which consists of programming noise, wear-out noise and retention noise. Model 2 is the complete channel model introduced in Sec. II-A.

*1) Model 1:* The channel model in [23] is an exponentially modified Gaussian distribution for each level. Cell-to-cell interference and programming error are not considered in this model. As a result, the channel characteristics of even and odd cells are the same. This channel model is still very similar to the multi-modal Gaussian model.

*2) Model 2:* The channel model proposed in Sec. II-A differentiates the even and odd cell read channels, and considers the effect of programming error and cell-to-cell interference.

## III. DYNAMIC VOLTAGE ALLOCATION WITH IDEAL CHANNEL INFORMATION

Dynamic Voltage Allocation (DVA) [22] is an algorithm that uses a single scaling factor $\alpha$ to attenuate the standard threshold voltages used to write to the cells. The DVA algorithm computes the appropriate scaling factor such that the read channel achieves at least a certain minimal value of mutual information required to support the rate at which information is stored (the rate indicated by the channel code). The range between the smallest and largest intended voltages is scaled so that it expands as the channel becomes more degraded until it reaches the range that would be employed in a system without DVA. In some Flash devices it may be possible to increase the range beyond the standard range without DVA, but that is not a focus of this paper.

For Flash memory described by Model 1, the mutual information is calculated as the difference of marginal and conditional differential entropies as follows [42]:

$$I(X;Y) = h(Y) - h(Y|X) . \quad (12)$$

Random variable $Y$ describes the measured threshold voltage, and $X$ represents the intended threshold voltage. The probability distribution of $Y$ is the channel distribution $f_Y(y)$, so

$$h(Y) = -\int_{-\infty}^{+\infty} f_Y(y) \log(f_Y(y)) \, dy . \quad (13)$$

The relevant conditional differential entropy is calculated as

$$h(Y|X) = -\int_{-\infty}^{+\infty} \sum_{i=1}^{4} f_{X,Y}(x_i, y) \log(f_{Y|X}(y|x_i)) \, dy , \quad (14)$$

Where the joint distribution is $f_{X,Y}(x,y) = f_{Y|X}(y|x)P_X(x)$, $P_X(x)$ is the probability mass function of $X$, and $f_{Y|X}(y|x)$ is the conditional PDF of $Y$ when the cell is written to a single level indicated by $x$. Let $\{v_i\}$ represent the set of default intended threshold voltages, the intended threshold voltage after scaling is $x_i = \alpha v_i$, where $\alpha$ is the scaling factor. As a result, adjusting the scaling factor will change both $h(Y)$ and $h(Y|X)$.

Because the cell-to-cell interference of a certain cell is a function of the threshold voltages of its surrounding cells, the Flash channel described by model 2 is a channel with memory. From [37], equation (12) is a lower bound on the actual mutual information because it treats interference as independent noise. Although equation (12) underestimates the mutual information, it is still a valid optimization objective function for DVA when cell-to-cell interference is treated as noise by the controller. If cell-to-cell interference is cancelled by signal processing in the controller, then the portion of cell-to-cell interference that is cancelled should be removed from the modeled noise before computing the mutual information.

As discussed in Sec. II, the channel probability distribution function is determined by the distributions of the five noise components. In general, the distribution can be calculated by convolving the five distributions. In [43], a model very similar to the one in Sec. II is presented and the analytical channel distribution function is calculated. The expression has a relatively high complexity. In this paper, numerical convolution is used to calculate an approximation of $f_{Y|X}(y|x)$. Note that for generating noise used in Monte Carlo simulation it is not necessary to compute this convolution. Rather, the individual noise terms, each of which is relatively simple in distribution, can be generated and added to the original signal.

Our implementation uses a bisection algorithm to find the scaling factor $\alpha$. The scaling factor has a range of [0,1], where a scaling factor of 1 corresponds to the maximum allowed threshold voltage for each level (the levels that would be used by a system not implementing DVA). There are two modes for the DVA algorithm. In the first mode, DVA makes adjustments based on the current channel distribution, and aligns the mutual information to a preset threshold that builds in a fixed margin to account for degradation before the scale factor is adjusted again. In the second mode, DVA makes adjustments based on the prediction of channel conditions in a future time point corresponding to the next DVA scale factor update. In this mode, if the channel model and channel parameter degradation models are correct, the mutual information at that future time point will be the target value. We use the first mode in this paper, because in practice the channel degradation models (especially the parameters in the models) may not accurately reflect the degradation process for device from different manufactures or for the same device under different working conditions (e.g. temperature).

Assuming ideal channel information, which is the exact model and parameters of the channel, a simulation can be conducted where the DVA algorithm adjusts the scale-factor every certain number of P/E cycles. In this paper, DVA functions every 100 P/E cycles for simplicity. The number of P/E cycles between DVA operations depends on how quickly the channel is varying and the appropriate interval itself might change over the lifetime of the device. For example, near the end of life when the channel is degrading rapidly a smaller interval might be appropriate.

In each iteration of the simulation, the DVA algorithm scales the intended threshold voltages to increase the channel mutual information to a predefined threshold. Because perfect knowledge of the channel state is assumed, the performance indicates the theoretical limit of the algorithm under these ideal conditions. For this ideal simulation we use Model 2, the full channel model presented in Sec. II.

DVA is most effective in extending lifetime when the major channel degradation is caused by the accumulated effect of charge traveling through the oxide layer of the Flash memory cells. Retention loss is the major degradation in this case. In this paper, the retention time is set to be a year to represent the worst possible channel for which a device is rated. If the actual retention time exceeds the fixed value, the channel could still provide enough mutual information depending on the distance between the actual channel mutual information and the mutual information limit for the channel code. Due to page limitations, DVA's performance when the retention time is 0 is not investigated. We believe the lifetime extension achieved by DVA will be limited in this case.

In this paper, all simulations assume MLC Flash (four levels) with each level used equally likely, and the target mutual information is 1.945 bits per cell for both even and odd channels. Because raw BER is determined by both the channel condition and the placement of read threshold voltages, there is no simple way to translate between mutual information and BER. However, simulation results show that a target of 1.945 bits achieves a raw BER of $10^{-2}$ in Fig. 8 using DTA to allocate read thresholds. We note that the raw BER of $10^{-2}$ is shown in [10] to be an operating point for Code 2 in [10] with 6 reads using a version of mutual-information-based DTA. In any case, the mutual information target can be adapted easily to whatever mutual information is needed to support a specific channel code (LDPC, BCH, or other). The target for the DVA algorithm is set to be 1.965 bits to provide an additional 0.02 bits of margin above the code-based target to allow for channel degradation before the next DVA update. The retention time is set to be one year. The default lifetime measured in P/E cycles for the programming error model is set to be 3000 P/E cycles.

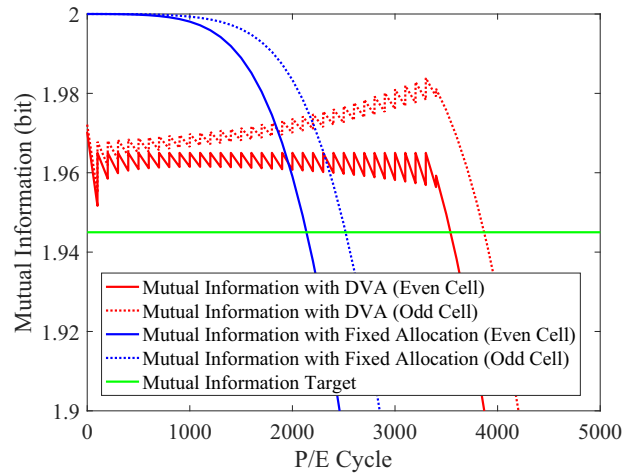

Fig. 4. DVA performance with ideal channel information v.s. fixed allocation's performance. (Ground truth channel is Model 2. DVA targets even channel. Channel parameters are listed in the appendix.)

The default intended threshold voltages are described by the vector $T = \begin{bmatrix} 2.8 & 5.2 & 6.4 & 7.86 \end{bmatrix}$ [22]. The actual threshold voltages resulting from DVA updates have the form $\alpha T$, and Fig. 6 shows how the scale factor $\alpha$ varies over time. When using fixed voltage allocation, the scaling factor $\alpha$ is fixed to be 1, so the default intended threshold voltages are used for each level through the entire lifetime of the device. When using DVA the scaling factor $\alpha$ starts at a value smaller than 1 and increases as the channel degrades until it reaches the value of 1 which corresponds to the maximum voltage levels supported by the device. The full set of channel parameters is given in the appendix.

For Flash memories using the even-odd write structure presented in Sec. II, the even-cell and odd-cell read channel characteristics are different as even cells are written before odd cells. The even cells experience more severe cell-to-cell interference and provide less mutual information under the same intended threshold voltage allocation. One approach to optimize to this type of Flash memory is to have a single DVA optimization process controlled by the mutual information of the even cells, which face the more severe channel conditions. This approach guarantees the worst case performance. Fig. 4 shows that with this approach the DVA algorithm can improve the lifetime of even cells by 65.7% from 2136 P/E cycles to 3540 P/E cycles and odd cells by 53.6% from 2517 P/E cycles to 3867 P/E cycles for the channel defined by the parameters in the appendix. Fig. 6 shows the scale factor $\alpha$ used for all cells (the black curve). Note the rapid reduction in the mutual information trajectory shown in Fig. 4 around the end of lifetime. This rapid reduction occurs after the scaling factor has become one and cannot be further increased as shown in Fig. 6.

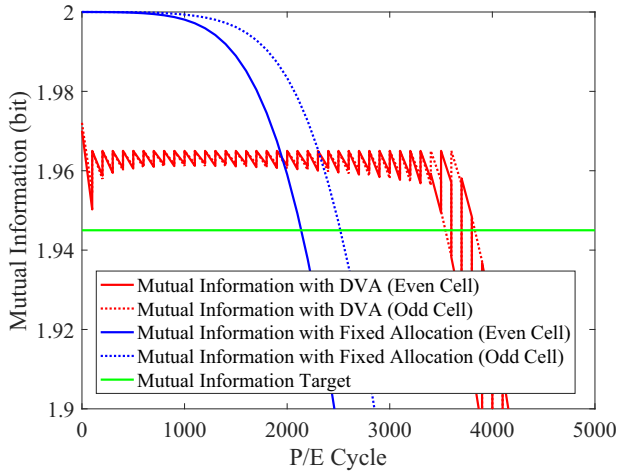

Fig. 5. DVA performance with ideal channel information and the performance of the fixed allocation. (The ground truth channel is Model 2. The writing order of pages is switched every 100 P/E cycles. Joint DVA is use to adjust both scaling factors for even and odd cells. Channel parameters are listed in the appendix.)

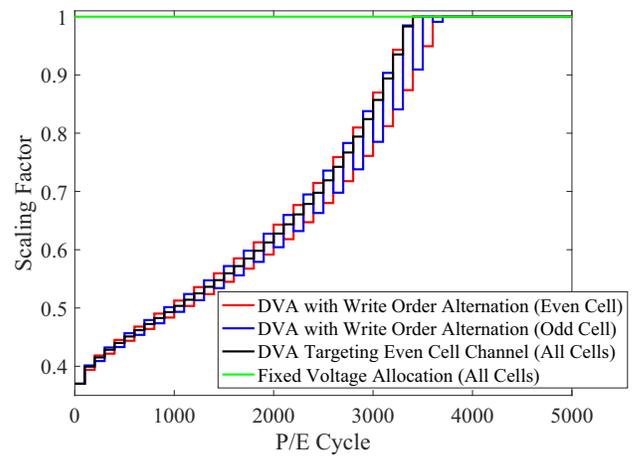

Fig. 6. Scaling factors generated by DVA with ideal channel information. (Result corresponds to Fig. 4 and 5. Ground truth model is Model 2. Channel parameters are listed in the appendix. The starting value of $\alpha$ is 0.37.)

One problem with the previous approach is that there is a significant performance difference between the even and odd cells. The intended threshold voltages provide too much mutual information margin for the odd cells. In fact, Fig. 4 shows the odd-cell mutual information margin increasing with time.

This situation is improved by using two separate DVA optimization processes in the system, one for the even-cell channel and one for the odd-cell channel. This solution will remove the excessive mutual information margin of the odd-cell channel, but the performance difference measured by lifetime will not be narrowed. One fundamental cause of the problem is that even cells experience more accumulated damage than odd cells.

To address this disparity, we alternate the writing order between even cells and odd cells every 100 P/E cycles. For example, even cells are written before odd cells from 0 to 99 P/E cycles, and odd cells are written before even cells from 100 P/E cycles to 199 P/E cycles. In this way, the two channels can experience similar accumulated damage by switching which channel receives the more severe cell-to-cell interference periodically. We apply DVA to this alternating-write-first approach with distinct scaling factors for the even-cell channel and the odd-cell channel, and use a joint DVA algorithm to adjust the two factors. The mutual information of the two channels will reach the predefined target together after each iteration. This approach equalizes the accumulated voltage $V_{acc}$ of even-cell and odd-cell channel, and narrows the difference of the two channels' characteristics. Fig. 5 shows that with this approach the DVA algorithm can improve the lifetime of even cells by 68.5% from 2136 P/E cycles to 3600 P/E cycles and odd cells by 40.7% from 2517 P/E cycles to 3542 P/E cycles for the channel defined by the parameters in the appendix. Fig. 6 shows the scale factor $\alpha$ used for even and odd cells as the red and blue curves respectively.

If the overall lifetime is defined as the number of P/E cycles until the mutual information of either channel crosses below the 1.945 bits target, joint DVA with the alternating-write-first approach extends the lifetime by 65.8% from 2136 P/E cycles to 3542 P/E cycles. It may seem surprising that the overall lifetime is improved by 65.8% when the lifetime of odd cells only improves by 40.7%, but the odd cells were lasting longer than the even cells to begin with so that the overall lifetime improvement is primarily driven by the improvement in the even cell lifetime.

Recall that the single DVA targeting even-cell channel extends the lifetime by 65.7% from 2136 P/E cycles to 3540 P/E cycles, so that the joint DVA did not improve lifetime significantly more than the simpler single DVA process for this channel model. When the strength factor of cell-to-cell interference $s$ increases, the average distortion of the interference for even-cell and odd-cell channels will rise, and the performance gain brought by the joint DVA approach will be more significant. For the remainder of this paper, the joint DVA with the alternating-write-first approach or its approximations will be used to implement DVA when the channel model in the simulation is Model 2, which includes cell-to-cell interference.

## IV. CHANNEL PARAMETER ESTIMATION

In order to dynamically allocate intended thresholds in a real system, the channel characteristics must be determined dynamically as they evolve. In [24], the authors demonstrate that accurate channel estimation can be achieved with limited computational complexity using multi-modal Gaussian channel model assumption and least squares algorithms. In [23], we further developed this approach with the relatively complex Model 1 channel model, which contains programming noise, wear-out noise and retention noise mentioned in Sec. II-A.

Using both the channel model and the parameter degradation model, the channel characteristics after a specific number of P/E cycles can be predicted for a specified retention time. However, the exact channel model and parameter degradation model need to be known in advance to enable this approach. When considering on-the-fly scenarios, we may also want to use actual measurements from the working device to determine the channel parameters without making detailed assumptions about the parameter degradation model or perhaps even the channel model itself.

In practical scenarios, empirical histograms of the threshold voltage can be measured with multiple read operations. Combined with the knowledge of the read reference voltages used in the measurements, the empirical distributions provide enough information about the ground truth channel (voltage) distribution. For a given channel model, this measured histogram can be used to estimate the model parameters using a least squares algorithm. This approach does not require the channel parameter degradation model.

In both [23] and [24], the channel model assumptions are constructed to be as close to the ground truth model as possible. In this paper, the idea of using a simple model assumption, different from the ground truth model, is explored.

### A. Channel Parameter Estimation Problem Formation

In this paper, channel parameter estimation is formulated as an optimization problem as demonstrated in [23]. One issue about this approach is that the formulation requires the calculation of derivatives of certain expressions. If the channel model is complex enough, there will be no analytical solution for the derivatives, and numerical approximation may need to be employed. As a result, the accuracy of the estimation could be affected. In both [23] and [24], the channel models facilitate the calculation of analytical derivatives.

### B. Binning Strategy and Least Squares Algorithms

In [23], we studied the binning strategy (the assignment of read threshold voltages that induce the number of bins and the bin placement of the resulting histogram) and the least square algorithm for histogram-based channel estimation for a Flash channel under Model 1. In [23], a 9-bin equal-probability binning strategy and Levenberg-Marquardt algorithm is demonstrated to have the optimal performance. This paper utilizes this result for all simulations.

Because DVA is an iterative process, at every iteration, bin placements are updated based on the channel estimation. The placements will only be optimal for the current channel condition even if the estimation is perfect. In order to provide more accurate placements for the next iteration, the placements are scaled by the ratio between the newly calculated scaling factor and the previous scaling factor used before the DVA process in this iteration. This approach provides a linear prediction of bin placements for next iteration's channel condition, and facilitates the channel estimation process.

## V. Dynamic Voltage Allocation with Model/Channel Mismatch

When the channel model agrees exactly with the actual channel, channel estimation can precisely characterize the channel if the estimated channel parameters are accurate. However, such perfect agreement of the channel model with the actual channel is hard to achieve in practice because of the many factors involved in shaping the Flash read channel characteristics. Even if the exact model can be known, the complexity of the model may preclude its use in channel estimation and DVA because of the associated complexity.

The lack of perfect agreement limits the precision with which the channel can be characterized even with the best-case parameter estimation. However, this problem can be controlled by selecting a reasonable *channel model assumption* that focuses on the most significant channel characteristics. A good choice can reduce the computational complexity of channel estimation and DVA while also capturing the essential behavior of the channel.

In [24], the authors investigate a scenario where the channel model assumption matches the actual channel and the channel model is a relatively simple multi-modal Gaussian distribution. In [23], we demonstrated the estimation performance where the channel model matched the actual channel exactly, but the more complex Model 1 of Sec. II-C was used. In both cases, high estimation accuracy is achieved. In this paper, DVA (and implicitly estimation) performance using the simple multi-modal Gaussian model as the assumption and the complex Models 1 and 2 of Sec. II-C for the actual channel is presented. Thus the channel parameters that need to be estimated for multi-modal Gaussian distribution are the variances and means of the distribution regardless of the numerous parameters used to create the actual channel used in the simulation.

### A. Model 1

Because this ground truth channel model is still very similar to the multi-modal Gaussian model, DVA can effectively extend Flash memory's lifetime although the channel model assumption is different than the ground truth. Fig. 7 shows the Monte Carlo simulation performance of DVA using a multi-modal Gaussian when Model 1 is the ground truth model. Here the lifetime is extended by 55.9% from 2683 P/E cycles to 4182 P/E cycles.

Note that unlike in Fig. 5, the *actual* mutual informations shown in Fig. 7 are not reset to exactly 1.965 bits after every DVA update. For the "Matching Assumption" case where the DVA channel model is the actual channel model, the variation in the mutual information after reset stems entirely from noise in the channel parameter estimation. For the "Gaussian Assumption" case, there is a large variation, which stems from parameter estimation error *and* from the fact that the DVA algorithm chooses the scale factor that sets the mutual information of the multi-modal Gaussian channel to 1.965 bits, but the mutual information of the actual channel turns out to be different. In particular, the mutual information of the actual channel is noticeably less than that of the multi-modal Gaussian towards the end of the device lifetime as the Gaussian assumption becomes less accurate. Still, the overall DVA approach is successful because the 0.02 bits of additional margin was sufficient to overcome the mismatch error.

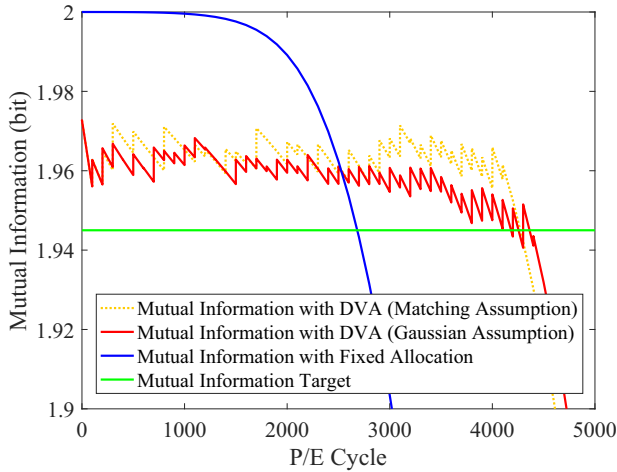

Fig. 7. DVA's performance with multi-modal Gaussian model. (Ground truth model is Model 1. Channel parameters are listed in the appendix.)

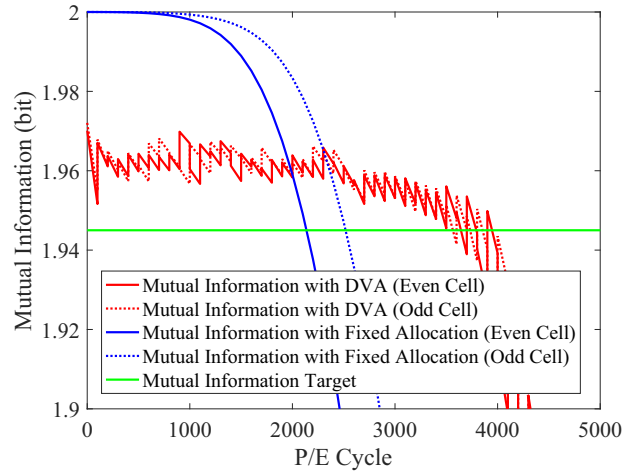

Fig. 8. DVA performance with multi-modal Gaussian model. The Ground truth model is Model 2. Channel parameters are listed in the appendix. The lifetime of even cells is extended by 70.5% from 2136 P/E cycles to 3642 P/E cycles. The lifetime of odd cells is extended by 41.6% from 2517 P/E cycles to 3564 P/E cycles. The overall lifetime is extended by 66.9% from 2136 P/E cycles to 3564 P/E cycles.

*B. Model 2*

For Model 2, the multi-modal Gaussian channel model assumption can only be considered as a rough approximation, but DVA can still provide a significant performance increase using this simple model. Fig. 8 shows the Monte Carlo simulation performance of DVA using the multi-modal channel model assumption while generating noise using Model 2. Because the only channel information available is the measured histograms from even and odd cells, DVA has to rely on channel estimation. The channel estimation only provides the estimated means and variances of the multi-modal Gaussian model rather than the parameters in the actual channel model.

The estimations of even-cell and odd-cell channels provide approximated channel characteristics for the odd and even channels respectively, when the write order is switched. Based on this observation, the scaling factors of the two channels can be updated based on the channel estimation of the other channel using DVA. This implementation is an approximation of the joint DVA approach in Sec. III. The simulation result shows that DVA can extend the overall lifetime by 66.9% from 2136 P/E cycles to 3564 P/E cycles for this channel.

Here, it is perhaps surprising that the overall lifetime is improved more by DVA employing the simple Gaussian model than by DVA using the more accurate model and prefect channel knowledge in Fig. 5. The Gaussian model underestimates the severity of the channel which causes it to use a smaller value of $\alpha$ that turns out to still provide sufficient mutual information. This extends lifetime a bit more by reducing the amount of charge written. Of course we could lower the amount of margin used in the simulation of Fig. 5 and also gain this benefit.

## VI. COMPARISON OF DYNAMIC VOLTAGE ALLOCATION AND DYNAMIC THRESHOLD ASSIGNMENT

Both dynamic voltage allocation and dynamic threshold assignment (DTA) [20], [21] track the degradation of Flash memory channel during its lifetime. DTA allocates appropriate *read* threshold voltages under different channel conditions to reduce the asymmetric errors caused by distribution shifting and widening. DVA allocates appropriate *write* threshold voltages under different channel conditions to adjust the mutual information to a sufficient (but not extravagant) level. DVA depends on the preset target mutual information which relates to the capability of error correction codes employed. Considering Flash memory as a M-ary baseband Pulse Amplitude Modulation (MPAM) communication system (4PAM in this paper), DTA provides hard decision thresholds closer to the optimal ones than fixed read thresholds, and DVA adjusts the average symbol power of the constellation to match channel conditions.

Assume the simplest fixed-read-voltage allocation schemes where hard decision boundaries are put in the midpoints between adjacent write threshold voltages, and DTA implementation from [21], simulations are conducted to compare the performance of DTA alone and DTA combined with DVA. Fig. 9 compares the raw BER (without error correction code) of reading with a fixed voltage allocation, DTA and DVA. The dash line in the figure represents the raw BER requirement for the LDPC code in [10] to function properly with soft information from 6 reads (Fig. 11 in [10]). To accommodate the rapid channel degradation at the beginning of the lifetime and meet the $10^{-2}$ raw BER target, the scaling factor for both even and odd cells are lower bounded by 0.45.

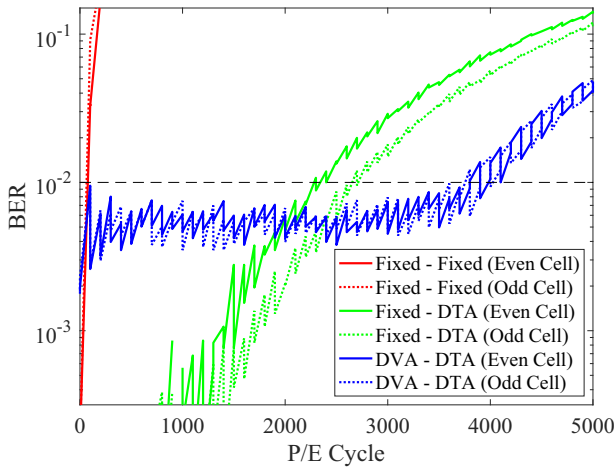

Fig. 9. DTA and DVA's performance. (Even and odd cells switch positions when using DVA. The legend follows the format *write voltage allocation algorithm - read voltage allocation algorithm*. Comparing the result from DVA - DTA with Fixed - DTA, The lifetime of even cells is extended by 66.5% from 2282 P/E cycles to 3800 P/E cycles. The lifetime of odd cells is extended by 41.4% from 2654 P/E cycles to 3754 P/E cycles. The overall lifetime is extended by 64.5% from 2282 P/E cycles to 3754 P/E cycles.)

Because the retention time is fixed to be one year in our simulations and there is no operation on the cells during this period, the stored voltage values in the cells suffer relatively significant shifts. As a result, fixed read voltage allocation provides very bad BER performance. With DTA, the read thresholds are able to track the channel degradation, thus produces BER performance consistent with the channel capacity using fixed write voltage allocation. DVA allocates write thresholds to maintain a reasonable channel capacity, so a non-zero low level raw BER appears from the beginning of the device's lifetime, and it is expected to be easily corrected by error correction codes. However, this is in contrast to DTA without DVA, where almost no errors occur in the first 1000 P/E cycles.

To achieve good raw BER performance, DVA requires that DTA-like algorithms are used in conjunction to provide decision boundaries that adapt to the changing write levels. Even without DVA, DTA-like algorithms are a practical necessity in order to track the movements of the channel distribution caused by channel degradations such as retention loss. Fig. 9 shows that DVA and DTA combined provide a significant performance improvement over DTA alone. The results indicate that DVA with DTA can extend the overall lifetime by 64.5% from 2282 P/E cycles to 3754 P/E cycles for this ground truth channel model compared with a fixed write-voltage allocation with DTA.

## VII. DYNAMIC VOLTAGE ALLOCATION WHEN VOLTAGE PLACEMENTS ARE QUANTIZED

In this section, we explore the performance of the DVA framework when voltage placements are limited to certain locations. In practice, hardware implementations introduce the constraint that both the read reference voltages and the intended write threshold voltage levels can only be set to certain values. This will add an additional constraint to both the intended threshold voltages and the read reference voltages, which affects DVA performance.

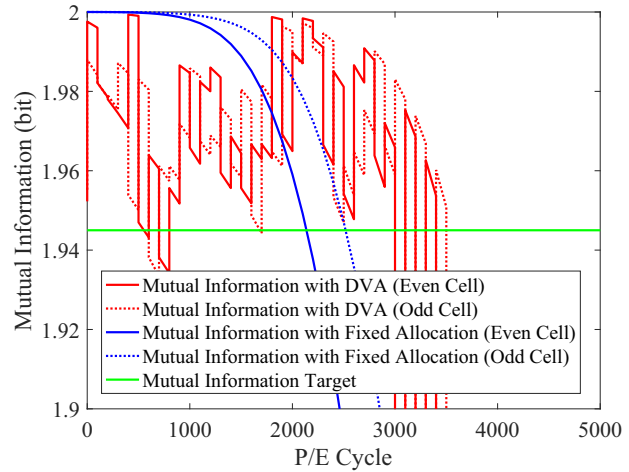

Fig. 10. DVA performance with quantized placements. (Quantization provides 64 possible locations. Ground truth model is Model 2. Channel parameters are listed in the appendix.)

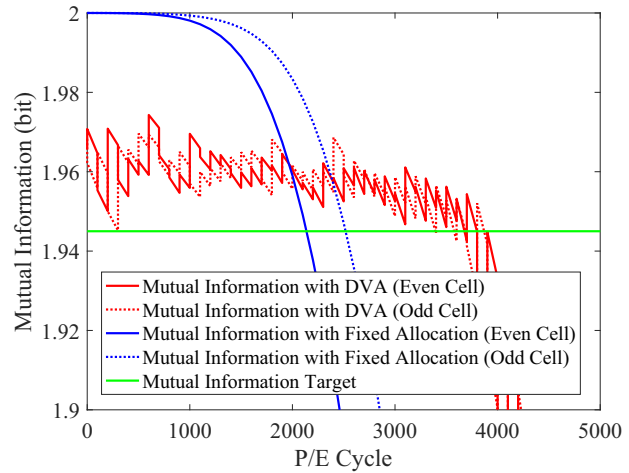

Fig. 11. DVA's performance with quantized placements. (Quantization provides 128 possible locations. Ground truth model is Model 2. Channel parameters are listed in the appendix. The lifetime of even cells is extended by 72.1% from 2136 P/E cycles to 3676 P/E cycles. The lifetime of odd cells is extended by 34.8% from 2517 P/E cycles to 3393 P/E cycles. The overall lifetime is extended by 58.8% from 2136 P/E cycles to 3393 P/E cycles.)

Monte Carlo simulation of DVA under several commonly used voltage quantization scenarios (64, 128 and 256-location quantization for the entire available voltage range) are conducted. We assume adjacent potential placement locations are separated by a constant distance, which corresponds to equal interval sampling over a certain voltage range. The quantization range is set to be from -1 Volt to 8 Volts. Figs. 10 and 11 show the mutual information v.s. P/E cycle curve when limiting the number of possible locations 64 and 128. Quantization with 256 possible locations results a performance very similar to Fig. 8. Quantization with 128 possible locations strikes a nice balance between DVA performance and the num-

TABLE I
COMPUTATIONAL COMPLEXITY OF READ THRESHOLD VOLTAGE ALLOCATION

| No. of ops. per iter. | ADD | DIV | LUT | No. of iterations |
|---|---|---|---|---|
| Boundary 1 (leftmost) | 9 | 4 | 4 | $\log_2 \frac{V_{r,max}-V_{r,min}}{\epsilon}$ |
| Boundary $i$, $2 \leq i \leq N-1$ | 9 | 4 | 4 | $\log_2 \frac{V_{r,max}-V_{r,i-1}}{\epsilon}$ |

Note: "ops." is the abbreviation for "operations". "iter." is the abbreviation for "iteration". The number of bins in the histogram is $N$. The $i$th boundary is $V_{r,i}$. The maximum and minimum possible read threshold voltage is $V_{r,max}$ ans $V_{r,min}$ respectively. The tolerance (error) of the bisection algorithm is $\epsilon$. ADD, DIV, LUT represent addition, division and table lookup operation respectively.

TABLE II
COMPUTATIONAL COMPLEXITY OF LEVENBERG-MARQUARDT ALGORITHM

|  | ADD | MUL | DIV | LUT |
|---|---|---|---|---|
| No. of ops. per iter. | $93N + \lceil \frac{N^3}{3} \rceil$ | $89N + \lceil \frac{N^3}{3} \rceil$ | $4N + \lceil \frac{N^3}{3} \rceil$ | $8N$ |

Note: "ops." is the abbreviation for "operations". "iter." is the abbreviation for "iteration". The number of bins in the histogram is $N$. The tolerance (error) of the bisection algorithm is $\epsilon$. ADD, MUL, DIV, LUT represent addition, multiplication, division and table lookup operation respectively.

TABLE III
COMPUTATIONAL COMPLEXITY OF SCALING FACTOR ADJUSTMENT

| ExADD | ExMUL | MI | No. of iterations |
|---|---|---|---|
| 5 | 4 | 1 | $\log_2 \frac{\alpha_{max}-\alpha_{min}}{\epsilon}$ |

(a) Number of operations per iteration

| ADD | MUL | DIV | LUT |
|---|---|---|---|
| $44n-1$ | $88n+1$ | $42n$ | $8n$ |

(b) Number of operations per MI

Note: The maximum and minimum possible scaling factor are $\alpha_{max}$ and $\alpha_{min}$ respectively. The tolerance (error) of the bisection algorithm is $\epsilon$. ExADD, ExMUL, MI represent extra addition, extra division in additional to the calculation of mutual information, and mutual information calculation respectively. The number of sample points used in MI calculation is $n$. ADD, MUL, DIV, LUT represent addition, multiplication, division and table lookup operation respectively.

ber of placement locations. When using 128-level quantization, the overall lifetime is extended by 58.8% from 2136 P/E cycles to 3393 P/E cycles.

Notice that the DVA framework often produces mutual information values that are too high (hence leading to a faster increase in $V_{acc}$ and faster degradation) when the number of possible placement locations is constrained to 64. We also observed that under this condition, the quantized read threshold voltage placements also suffered. Even though only nine reads are performed, with only 64 possible values, two or more would be in the same place. As a result, the effective resolution of the measured histogram is reduced, and the channel parameter estimation algorithm becomes less accurate and degrades DVA performance.

## VIII. COMPLEXITY ANALYSIS

The DVA framework consists of two processes: channel parameter estimation and scaling factor adjustment. This section considers complexity for the case discussed in Sec. V where we use a multi-modal Gaussian distribution as the channel model for both channel parameter estimation and scaling factor adjustment. For the channel parameter estimation process, there are two steps: read threshold voltage assignment and iterative Levenberg-Marquardt parameter estimation. The computational complexity of DVA can be analyzed according to these three primary components: read threshold voltage assignment, iterative Levenberg-Marquardt parameter estimation, and scaling factor adjustment.

Read threshold voltage assignment is used to provide boundaries for the bins in the histogram satisfying the equal-probability binning strategy discussed in Sec. IV-B. In the DVA framework, this assignment is calculated using a Gaussian channel model and the estimated channel parameters (means and variances) from the parameter estimation process 100 P/E cycles ago. Integration of a Gaussian probability distribution function is needed, but the operation is in essence the evaluation of the Q-function (tail probability of the standard normal distribution), and can be implemented with a look-up table. The exact position of each read is calculated using bisection method. Table I summarizes the number of additions (ADD), divisions (DIV) and table lookups (LUT) in each iteration, and the number of iterations required to calculate each read threshold voltage.

The computational complexity of using the Levenberg-Marquardt algorithm based on the Gaussian channel model assumption to estimate the Flash read channel is carefully analyzed in [24]. The results are summarized here in Table II. Note that the $\lceil \frac{N^3}{3} \rceil$ term in the table comes from the inversion operation of an $N \times N$ matrix using the Gauss-Jordan method, where $N$ is the number of bins in the measured histogram. Because the Levenberg-Marquardt algorithm has multiple stopping criteria, the exact number of iterations needed is not a fixed value for different channel conditions. In our simulations, we observe that the maximum number of iterations is less than 100.

The scaling factor adjustment process uses a bisection algorithm to search for the smallest scaling factor that achieves the mutual information target given the estimated means and variances. Table III(a) summarizes the number of extra additions (ExADD), multiplications (ExMUL) other than the ones in the mutual information calculation, the number of mutual information calculations (MI) in each iteration, and the number of iterations required to calculate the scaling factor. The most computation-heavy operation in this process is the calculation of the mutual information (MI) of a multi-modal Gaussian distribution, which involves integration operations. Assuming the Gauss-Hermite quadrature [44] method is used to approximate the integration, the mutual information calculation needs to be represented in the form of

$$\int_{-\infty}^{\infty} e^{-x^2} f(x) \mathrm{d}x \ , \qquad (15)$$

where $f(x)$ is a function. Then the following approximation

can be used:

$$\int_{-\infty}^{\infty} e^{-x^2} f(x) \mathrm{d}x \approx \sum_{i=1}^{n} w_i f(x_i) , \quad (16)$$

where $x_i$'s are the roots of the Hermite polynomial $H_n(x)$, $w_i$'s are the weights calculated by

$$\frac{2^{n-1} n! \sqrt{\pi}}{n^2 [H_{n-1}(x_i)]^2} , \quad (17)$$

and $n$ is the number of sample points. The accuracy of the approximation increases with $n$. Following (12-14), the new mutual information representation is

$$I(X,Y) = -\frac{1}{4\sqrt{\pi}} \sum_{i=1}^{4} \int_{-\infty}^{\infty} e^{-z_i^2} \log(g(z_i)) \mathrm{d}z_i , \quad (18)$$

where

$$g(z_i) = \frac{1}{4} + \sum_{\substack{j=1 \\ j \neq i}}^{4} \frac{\sigma_i}{4\sigma_j} e^{z_i^2 - \frac{(\sqrt{2}\sigma_i z_i + x_i - x_j)^2}{2\sigma_j^2}} . \quad (19)$$

Assuming all the weights can be pre-calculated, and $\log(\cdot), e^{(\cdot)}$ can be calculated with lookup tables, the number of operations needed for each MI calculation is shown is Table III(b).

## IX. Conclusion

This paper introduces a framework to extend Flash memory lifetime by dynamic allocation of intended threshold voltage levels based on the current mutual information of the Flash read channel. Analysis and simulation results demonstrate that this framework can provide significant lifetime extension in practical settings. Simple channel models can be used to estimate a complex Flash channel, and optimize the voltage placements. Good performance can be achieved even when the placement of voltages is constrained to 128 possible values. The frame can be modified to reduce computational complexity while providing comparable performance to a fully implemented system.

The results in this paper can be improved by extending the ground truth channel models to include additional mechanisms for channel degradation in Flash memory and by using data measured from actual Flash devices. Another direction to further develop the DVA framework is to explore using raw BER as the optimization target for the DVA framework.

## Appendix
### Channel Parameters used in This Paper

| Programming Noise [22] | | | |
|---|---|---|---|
| $\sigma_e$ | 0.35 | $\sigma_p$ | 0.05 |
| Wear-out Noise [22] | | | |
| $A_w$ | $1.8 \times 10^{-4}$ | $C_w$ | $1.26 \times 10^{-3}$ |
| $k_i$ | 0.62 | —— | —— |
| Retention Noise [22] | | | |
| $A_r$ | $7.0 \times 10^{-4}$ | $B_r$ | $4.76 \times 10^{-3}$ |
| $k_i$ | 0.62 | $k_o$ | 0.3 |
| $t_0$ | 1(hour) | $t$ | 8760 |
| Cell-to-cell Interference [39] | | | |
| $E\{\gamma_x\}$ | $0.1s$ | $E\{\gamma_y\}$ | $0.08s$ |
| $E\{\gamma_{xy}\}$ | $0.006s$ | $s$ | 0.2 |
| $w_k$ | $0.2E\{\gamma_k\}$ | $\sigma_k$ | $0.3E\{\gamma_k\}$ |
| Programming Error [38] | | | |
| $P_{0,2}$ | exp(0.87x-11.89) | $P_{0,3}$ | exp(1.41x-19.82) |
| $P_{1,2}$ | exp(1.63x-19.22) | $P_{1,3}$ | exp(0.73x-11.67) |
| $P_{2,3}$ | exp(1.50x-17.69) | —— | —— |

Default intended threshold voltages [22]: $[2.8, 5.2, 6.4, 7.86]$. $V_{max} = 16.$ [22]